\documentclass[a4paper,twocolumn,superscriptaddress,10pt, allowfontchageintitle,accepted=2017-11-06]{quantumarticle}
\pdfoutput=1
\usepackage[utf8]{inputenc}
\usepackage[english]{babel}
\usepackage[T1]{fontenc}
\usepackage{amsmath}
\usepackage{hyperref}
\hypersetup{colorlinks=true,linkcolor=red,citecolor=blue}

\usepackage[numbers,sort&compress]{natbib}
\usepackage{tikz}

\usepackage{amssymb}
\usepackage{amsthm}
\usepackage{graphicx}
\usepackage{color}
\usepackage{psfrag}
\usepackage{amsfonts}
\usepackage{multirow}

\newcommand{\ket}[1] {| #1 \rangle}

\newcommand{\mean}[1]{\left\langle #1 \right\rangle}

\newcommand{\one}{\leavevmode\hbox{\small1\normalsize\kern-.33em1}}

\DeclareMathOperator{\tr}{tr}

\begin{document}

\title{{\fontsize{23}{23}\selectfont Probing the Non-Classicality of Temporal Correlations}}

\author{Martin Ringbauer}
\affiliation{Centre for Engineered Quantum Systems, School of Mathematics and Physics, University of Queensland, Brisbane, QLD 4072, Australia}
\affiliation{Centre for Quantum Computer and Communication Technology, School of Mathematics and Physics, University of Queensland, Brisbane, QLD 4072, Australia}
\affiliation{Institute of Photonics and Quantum Sciences, School of Engineering and Physical Sciences, Heriot-Watt University, Edinburgh EH14 4AS, UK}
\orcid{0000-0001-5055-6240}

\author{Rafael Chaves}
\affiliation{International Institute of Physics, Federal University of Rio Grande do Norte, 59070-405 Natal, Brazil}
\orcid{0000-0001-8493-4019}

\date{\today}

\begin{abstract}
Correlations between spacelike separated measurements on entangled quantum systems are stronger than any classical correlations and are at the heart of numerous quantum technologies. In practice, however, spacelike separation is often not guaranteed and we typically face situations where measurements have an underlying time order. Here we aim to provide a fair comparison of classical and quantum models of temporal correlations on a single particle, as well as timelike-separated correlations on multiple particles. We use a causal modeling approach to show, in theory and experiment, that quantum correlations outperform their classical counterpart when allowed equal, but limited communication resources. This provides a clearer picture of the role of quantum correlations in timelike separated scenarios, which play an important role in foundational and practical aspects of quantum information processing.
\end{abstract}

\maketitle

Quantum predictions are fundamentally incompatible with the intuitive notion of cause and effect that underpins all of classical empirical science, as well as everyday experience. Most famously, Bell's theorem~\cite{Bell1964} shows that the correlations revealed by measurements on distant parts of entangled quantum systems cannot be explained in causal terms, under the assumptions that local measurement outcomes are not influenced by spacelike separated events---\emph{local causality}---and that measurement settings can be chosen freely. This phenomenon has become known as quantum nonlocality, and has been extensively studied for variants of the scenario above, where spacelike separated parties perform measurements on a composite quantum system~\cite{Brunner2014}. 

Yet, in practice we often face situations where spacelike separation between observers is not guaranteed and instead there is some time order underlying the observed physical events. Indeed, it has been shown that a single quantum system measured at different points in time exhibits Bell-like correlations~\cite{Brukner2004,Markiewicz2014,Budroni2013,Fritz2010,Fedrizzi2011,Pawlowski2010a}. However, in such a scenario one cannot appeal to relativity to rule out communication between the time steps. This can be problematic, since a classical model with sufficient communication can simulate any correlations. Recent research has thus been devoted to clarifying to what extent the incompatibility between classical and quantum descriptions encountered in the spacelike case carries over to the timelike scenario. 

Specifically, one research direction has been to study how limited resources such as entropy~\cite{Chaves2015b}, memory~\cite{Galvao2003,Montina2008,Zukowski2014} or dimension~\cite{Gallego2010,Brierley2015,Bowles2015} can lead to a quantum advantage over classical models. On the other hand, one can relax the local causality assumption in Bell's theorem, aiming to explain quantum correlations by classical models augmented with communication~\cite{Bacon2003,Toner2003,Pironio2003,Vertesi2009,Hall2011,Maxwell2014,Chaves2015,Brask2016,Ringbauer2016,Chaves2016causal}. However, communication is typically only allowed for the classical system, leading to an unfair comparison between classical and quantum models, and it remains unclear to what extent these results hold when equal communication power is given to quantum mechanics~\cite{Brask2016}.

Here we use a causal modeling approach to allow a fair comparison of classical and quantum models of timelike separated correlations. First, we show that for two time-ordered spatially separated measurements, augmented with a limited amount of classical communication, quantum correlations outperform their classical counterpart. Second, we show that, in contrast to spatial Bell-type scenarios~\cite{Wood2015}, there are faithful classical causal models reproducing all the temporal correlations obtained from a series of projective quantum measurements on a single quantum system. However, we find that non-classical correlations can arise in this scenario, when considering a slightly weaker classical causal model, that is nonetheless strictly stronger than previous results~\cite{Leggett1985}, which are contained as a special case. Finally we derive Bell-type inequalities for the above scenarios and demonstrate that they are violated in a photonic experiment.

\text{ }\\
\text{ }\\
\text{ }\\

\section{Causal modeling and timelike Bell-scenarios}
In the following we employ the formalism of Bayesian networks~\cite{Pearlbook}, which provides a natural framework for classical causal modeling. A central concept in this framework is that of a \emph{directed acyclic graph} (DAG), which consists of a set of nodes, representing the relevant random variables\footnote{We adopt the standard convention that uppercase letters label random variables while their values are denoted in lower case.} in the considered situation, and directed edges, representing the causal relations between those variables. A set of variables $X_1,\dots,X_n$ forms a Bayesian network with respect to some DAG if and only if the probability distribution  $p(x_1, \dots , x_n)$ can be decomposed as
\begin{equation}
p(x_1, \dots , x_n)=\prod_{i=1}^{n} p(x_i \vert pa_i)
\end{equation}
where $PA_i$ stands for the set of graph-theoretical parents of the variable $X_i$ (i.e.\ all variables that have a direct causal influence over $X_i$). Without loss of generality each variable can be understood as a deterministic function of its parents plus local noise $U_i$ that supplies potential randomness, $x_i=f_i(pa_i,u_i)$. This formalism thus enables a distinction between simple statistical correlations and actual causation by explicitly specifying the underlying mechanism generating the data.

Here we are interested in DAGs containing so-called \emph{latent variables}, which are empirically inaccessible. In the context of Bell's theorem~\cite{Bell1964} these are also known as \emph{hidden variables}. For any set of observed correlations, there are in general many DAGs with hidden variables that could have produced these observations. Among these, causal inference is particularly interested in those fulfilling the conditions of \emph{minimality} and \emph{faithfulness}. Minimality requires that, given two possible causal models, we choose the simplest one, capable of generating the smallest set of correlations (including the observed one). In turn, faithfulness, requires the causal model to be able to explain the observed data without resorting to fine-tuning of the causal-statistical parameters. In other words, any observed (conditional) independence should be a consequence of the causal structure itself, rather than a specific choice of parameters. Faithful (i.e.\ non-fine-tuned) models are therefore robust against changes in the causal parameters and thus the preferred choice.

\begin{figure}[h!]
\begin{center}
\includegraphics[width=0.8\columnwidth]{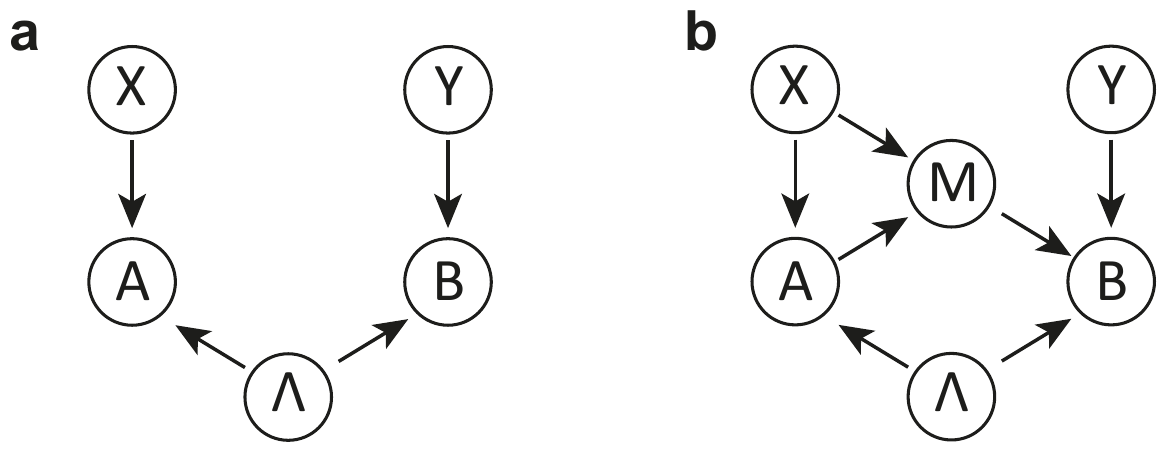}
  \end{center}
\vspace{-1.5em}
\caption{\textbf{Causal structure underlying Bell's theorem.} \textbf{(a)} Two observers, Alice and Bob, each have the choice of two measurements represented by the random variables $X$ and $Y$, respectively. The correlation between their measurement outcomes, modeled as random variables $A$ and $B$, respectively, are mediated solely by a common cause in their past---the hidden variable $\Lambda$. \textbf{(b)} Bell's causal model augmented with one-way communication from Alice to Bob. The initial state of the joint system is specified by the ontic state $\Lambda$. First, Alice performs a measurement with setting $x$, obtaining outcome $a$. She then sends a message $m$ to Bob, who performs a measurement with setting $y$, obtaining outcome $b$.}
\label{fig.BellDag}
\end{figure}

To illustrate the last point, consider the paradigmatic causal structure of Bell's theorem in Fig. \ref{fig.BellDag}a. This structure intuitively reflects the causal assumptions of Bell's theorem, leading to the so-called local hidden-variable (LHV) models. First, the two parties are assumed to be spacelike separated, such that the correlations between the measurements outcomes $A$ and $B$ can only be mediated via a common source $\Lambda$, implying that $p(a,b\vert x,y,\lambda)= p(a \vert x, \lambda)p(b \vert y, \lambda)$. Second, it is assumed that the experimenters can freely choose which observables to measure (represented by the random variables $X$ and $Y$), independently of how the system was prepared, that is, $p(x,y,\lambda)=p(x,y)p(\lambda)$. Note that these constraints implied by the causal model appear at an unobservable level since they explicitly involve the hidden variables $\Lambda$. Yet, they also imply observable constraints in the form of no-signaling conditions, expressed as $p(a \vert x,y)=p(a \vert x)$ and $p(b \vert x,y)=p(b \vert y)$, and Bell inequalities~\cite{Bell1964,Bell1976}.

While quantum correlations obey the no-signaling conditions, they violate Bell inequalities~\cite{Bell1964,Bell1976} and are thus in conflict with the assumptions behind the causal structure in Fig.~\ref{fig.BellDag}a. In order to maintain a classical causal explanation, the model in Fig.~\ref{fig.BellDag}a must therefore be augmented with additional resources; something that can only be done at the cost of introducing fine-tuning~\cite{Wood2015}. For instance, the causal structure in Fig.~\ref{fig.BellDag}b can reproduce all quantum correlations, but at the same time allows, in principle, for non-local correlations between $X$ and $B$. Hence, in order to satisfy the no-signaling condition $p(b \vert x,y)=p(b \vert y)$ the causal parameters must be chosen from a set of measure zero~\cite{Pearlbook}, a signature of fine-tuning.

Studying such non-local classical models can provide valuable insights into the relation between classical and quantum theory, and their applications~\cite{Brunner2014}. However, at the same time such models lead to an unfair comparison, since allowing for communication makes not only classical, but also quantum models more powerful. In practice, it is more natural to assume a certain underlying causal structure, and ask what can be achieved with classical and quantum resources? Bell's theorem is a particular case of this broader question, referring to spacelike separated events. However, there are often situations where the events are timelike rather than spacelike ordered. Examples include central quantum information tasks, such as teleportation \cite{Bennett1993}, superdense coding \cite{Bennett1992}, and measurement-based quantum computation \cite{Briegel2009}, as well as prepare-and-measure scenarios \cite{Gallego2010}, sequential Bell scenarios \cite{Popescu1995,Gallego2014}  and a sequence of measurements on a single quantum system \cite{Leggett1985}.

\section{Non-classicality of timelike correlations augmented by communication}
Consider the scenario in Fig.~\ref{fig.BellDag}b, where two distant parties, Alice and Bob, share pre-established correlations (represented by $\Lambda$) and are allowed one-way communication (the message $M$). As shown in Ref.~\cite{Bacon2003}, a classical model of this form (for a large enough message $M$) is enough to reproduce all the correlations obtainable from local measurements on two-qubit entangled states as in Fig.~\ref{fig.BellDag}a, which are described by
\begin{equation}
p(a,b \vert x,y)= \mathrm{Tr} \left[ (M^a_x \otimes M^b_y) \rho \right] ,
\end{equation}
where $M^a_x$ and $M^b_y$ are measurement operators for Alice and Bob, respectively, and $\rho$ is the density matrix describing the shared quantum state. If, in contrast, we impose the causal structure of Fig.~\ref{fig.BellDag}b also to the quantum case---that is, Bob's measurement may depend on Alice's measurement setting and outcome---then the set of correlations is described by
\begin{equation}
p(a,b \vert x,y)= \sum_{m}\mathrm{Tr} \left[ (M^a_x \otimes M^b_{y,m}) \rho \right] .
\label{eq:quantum_oneway}
\end{equation}
Note that Bob's measurement operator now explicitly depends on the values of $X$ and $A$ (via the message $M$) \footnote{In fact, Bob could apply to his share of the joint quantum state any completely-positive trace-preserving map dependent on the message $m$.}. Clearly, if there are no restrictions on the dimensionality of the message, every distribution of the form above can be also obtained by the classical hidden variable model in Fig.~\ref{fig.BellDag}b, that is, by a model respecting the decomposition
\begin{equation}
p(a,b \vert x,y)= \sum_{m,\lambda} p(\lambda)p(m \vert x,a) p(a\vert x,\lambda) p(b\vert y,m,\lambda).
\end{equation}
In fact, it is enough to choose $m=x$ to reproduce all possible one-way signalling correlations. To see this, note that the quantum correlations arising from Eq.~\eqref{eq:quantum_oneway} are of the form $p(a,b \vert x,y)=p(b \vert x,y,a)p(a \vert x)$, and that $a$ can be made a deterministic function $a=f_a(x,\lambda)$. Hence $a$ does not carry any information that is not already contained in $\lambda$ and $x$.

Notwithstanding, this picture changes if we impose restrictions on the message sent from Alice to Bob. Consider that each party measures three dichotomic observables (i.e.\ $x,y=0,1,2$ and $a,b=0,1$) and that Alice is bound to send a binary message ($m=0,1$). In this case, every classical model must obey the inequality~\cite{Bacon2003}
\begin{align}
\label{eq.M332}
 \mathcal{S}_\mathrm{1 bit}=& \mean{A_0B_0}+\mean{A_0B_1}+\mean{A_0B_2}+\mean{A_1B_0} \\ \nonumber
& +\mean{A_1B_1}-\mean{A_1B_2}+\mean{A_2B_0}-\mean{A_2B_1} \leq 6 .
\end{align}
Furthermore, it was shown in Ref.~\cite{Bacon2003} that this inequality also holds for correlations from local measurements on entangled quantum states, while it can be violated by more powerful no-signalling correlations. Hence, while one bit of communication is in this scenario sufficient for a classical model to simulate quantum correlations (without communication), it is not sufficient to simulate all possible no-signalling correlations. However, just like communication-augmented classical models become more powerful, so do quantum models. Specifically, local measurements on entangled quantum states augmented with one bit of communication can indeed violate inequality~\eqref{eq.M332}~\cite{Brask2016}, thus showing that under fair comparison, quantum advantage persists in such a timelike-separated Bell scenario.

\begin{figure}[h!]
  \begin{center}
 \includegraphics[width=\columnwidth]{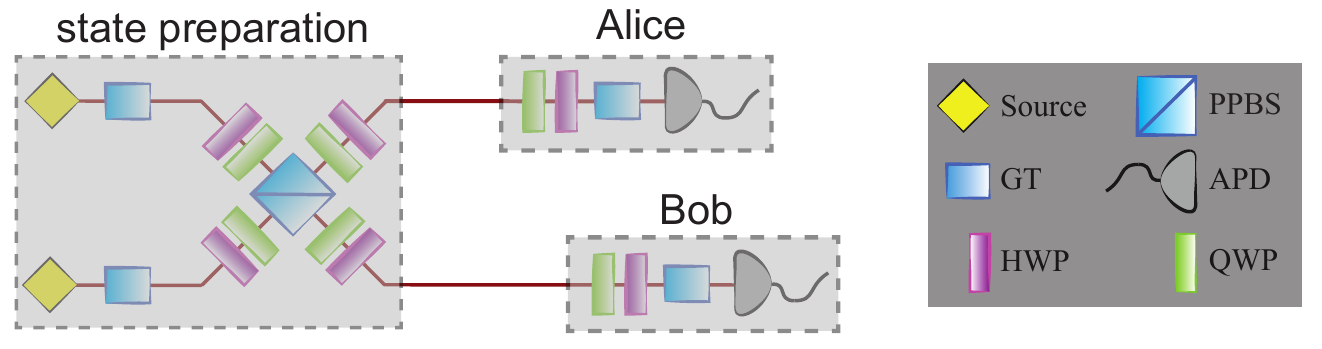}
  \end{center}
\vspace{-1.5em}
\caption{\textbf{Experimental setup for studying the non-classicality of timelike correlations}. Pairs of single photons are produced via spontaneous parametric downconversion in a $\beta$-barium borate crystal (not shown). The two photons are entangled to arbitrary degree using a non-deterministic controlled-NOT gate (CNOT), based on nonclassical interference in a partially polarizing beam splitter (PPBS)~\cite{Langford2005}. Alice and Bob then perform local projective measurements on their share of the entangled state, which are implemented using a set of half- (HWP) and quarter-waveplates (QWP), a Glan-Taylor polarizer (GT) and single-photon counters (APD).}
  \label{fig:Setup}
\end{figure}

As an example, consider that Alice and Bob share a maximally entangled state $\ket{\Psi^{+}}=(\ket{00}+\ket{11})/\sqrt{2})$. Alice performs local measurements with settings $A_0=A_1=\hat X$, $A_2=\hat Z$ and encodes her measurement setting in a message $m$ to Bob. For $x=0$ she sends $m=0$ and for $x=1$ or $x=2$ she sends $m=1$. Assuming that all inputs are equally likely this message has an entropy of $H(m) \sim 0.92$ and thus contains less than 1 bit of information. If Bob receives $m=0$, he measures $B_0=B_1=B_2=\hat X$, while for $m=1$ he measures $B_0=(\hat X+\hat Z)/\sqrt{2}$, $B_1=(\hat X-\hat Z)/\sqrt{2}$ and $B_2= - \hat X$. This protocol achieves $\mathcal{S}_\mathrm{1 bit}=4\sqrt{2}$, thus violating inequality~\eqref{eq.M332}. 

\begin{figure}[h!]
  \begin{center}
\includegraphics[width=0.95\columnwidth]{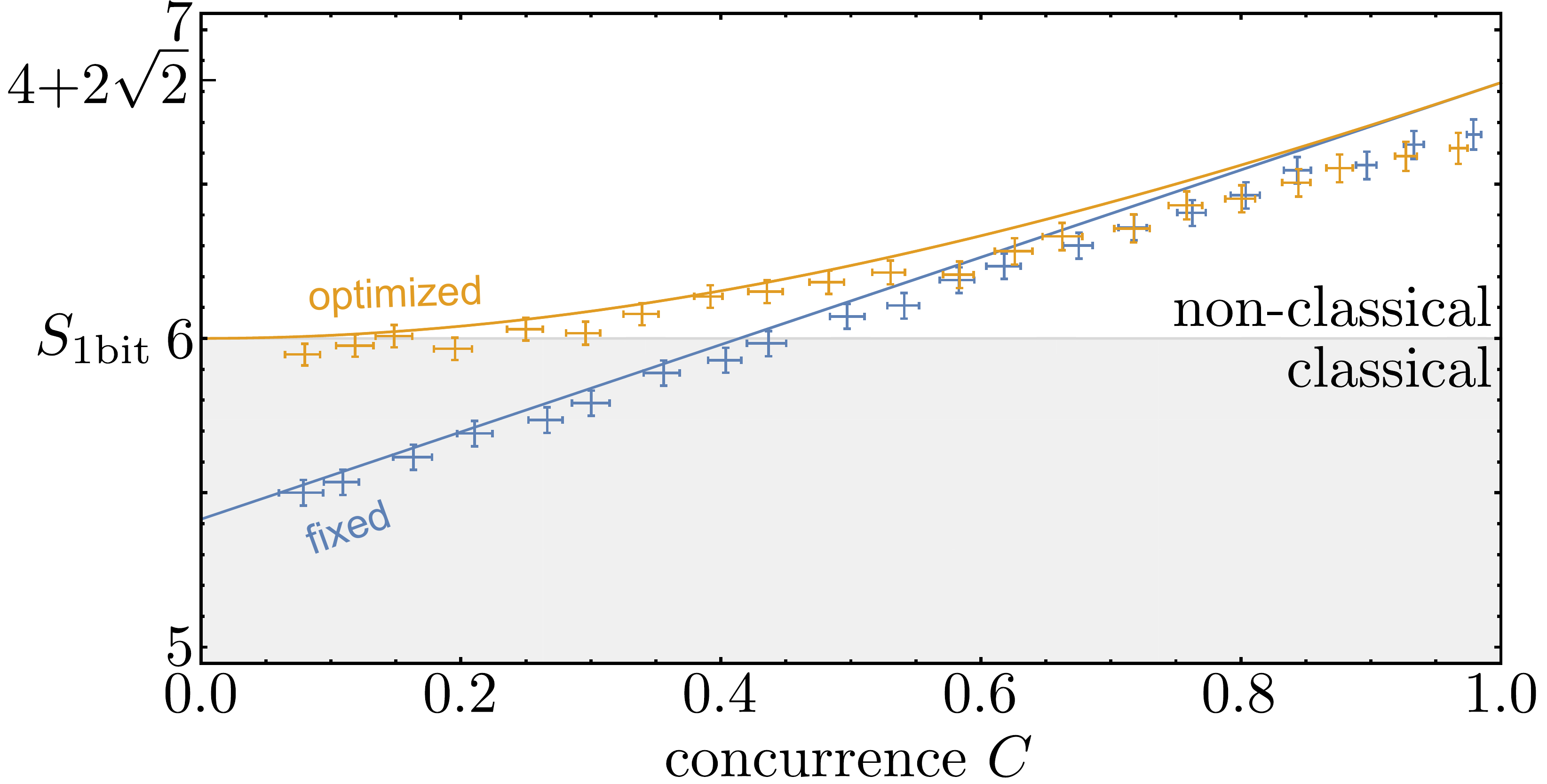}
  \end{center}
\vspace{-1.5em}
\caption{\textbf{Experimental test of inequality~\eqref{eq.M332} for temporally-separated spatial correlations}. The inequality is tested for a family of states with varying degree of entanglement as quantified by the concurrence. Data points include $3\sigma$ statistical error bars and the theory prediction is shown as solid lines. Data points in the grey region are compatible with inequality~\eqref{eq.M332}, while it is violated for data in the white region. Blue data and theory corresponds to the fixed measurement scheme outlined in the text, whereas the orange data and theory is obtained when the measurement settings are optimized for a given non-maximally entangled state, see Appendix~\ref{Sec:Supp2} for more details.}
  \label{fig:Results2}
\end{figure}

Experimentally we can test inequality~\eqref{eq.M332} with qubits encoded in the polarization of single photons, see Fig.~\ref{fig:Setup}. Two single photons are first entangled using a non-deterministic controlled-NOT gate, and then distributed to Alice and Bob. By varying the input states this configuration can produce states with arbitrary degree of entanglement, quantified by the concurrence $\mathcal{C}$~\cite{Hill1997}, see Appendix~\ref{Sec:Supp2} for more details. For simplicity, the message $m$ has been directly taken into account in Bob's measurement basis. The experimental results in Fig.~\ref{fig:Results2} show a clear violation of inequality~\eqref{eq.M332}, up to $S_{\text{1bit}}=6.66^{+0.02}_{-0.02}$.

\section{Classical causal models for a sequence of projective measurements}
Above we have analyzed the correlations between two timelike separated parties, whose measurements follow an underlying time order. We will now focus on a series of projective measurements performed on a single quantum system.
A measurement with setting $x$, obtaining outcome $a$, is described by a set of projective operators $M^{x}_{a}$, such that after the measurement the system, initially in state $\rho_1$, is left in a state given by $\rho_2=\rho^{x}_{a}= M^{x}_{a}$. In a next time step a measurement with setting $y$, producing outcome $b$ is performed, leaving the system in a state $\rho_3=\rho^{y}_{b}=M^{y}_{b}$, and so forth.

\begin{figure}[t!]
  \begin{center}
\includegraphics[width=0.9\columnwidth]{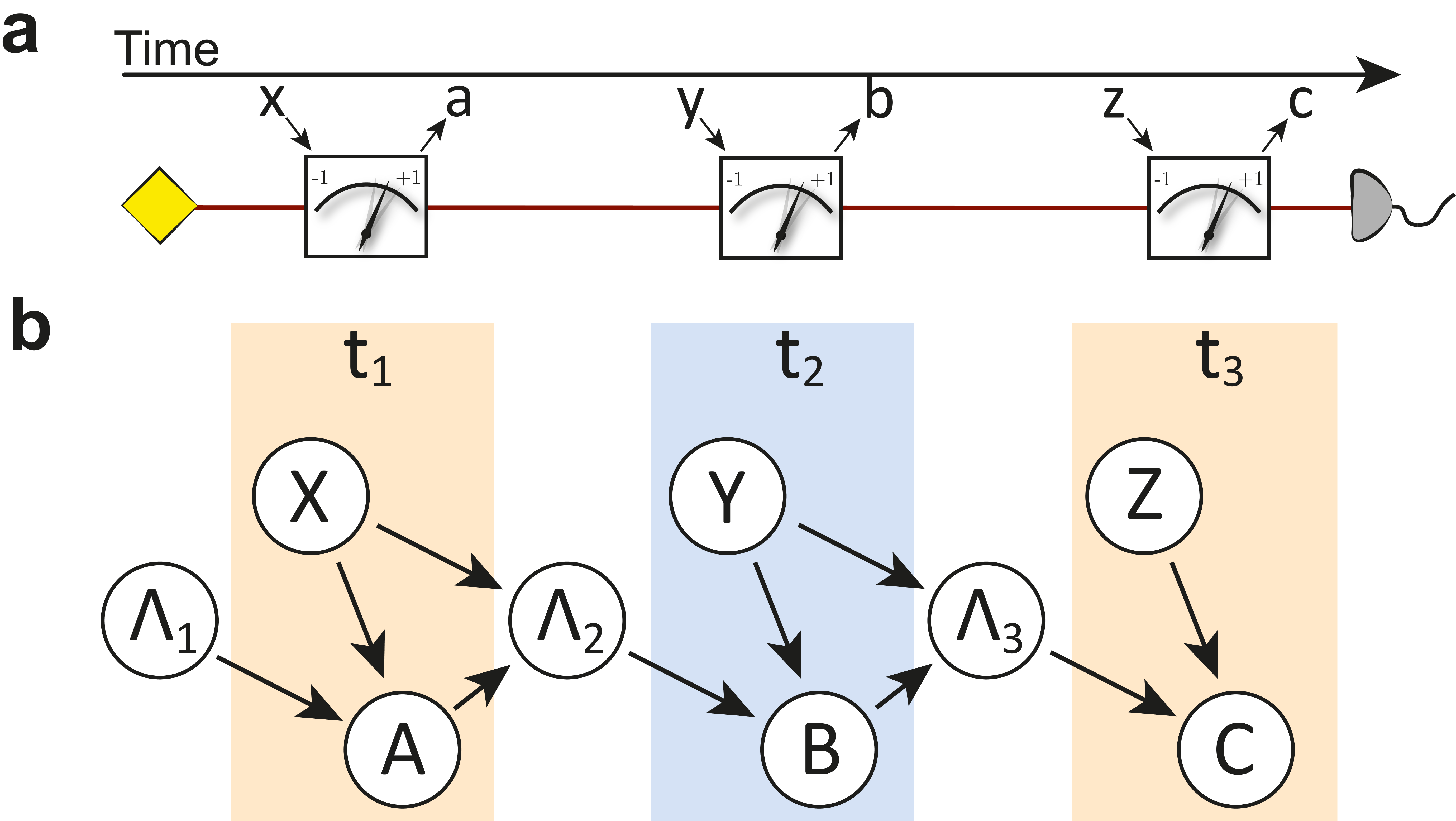}
  \end{center}
\vspace{-1em}
\caption{\textbf{A sequence of three projective measurement on a single qubit.} \textbf{(a)} A single quantum system is subject to a sequence of projective measurements at times $t_1$, $t_2$, and $t_3$ with settings $x,y,z$, and obtaining outcomes $a,b,c$, respectively. \textbf{(b)} A general classical causal model for the scenario in $(a)$. Note that, although causal graphs are formulated without reference to any spacetime structure, here we have drawn the graph such that the horizontal direction can be identified with time.}
\label{fig.dag1}
\end{figure}

Similar to a standard Bell experiment, the classical causal description of this scenario, illustrated in Fig.~\ref{fig.dag1}, involves a random variable $\Lambda_1$---the \emph{ontic state}~\cite{Harrigan2010}---which fully specifies the initial state of the system. The probability that a measurement $x$ produces the outcome $a$ is then given by $p(a \vert x)= \sum_{\lambda_1} p(a\vert x,\lambda_1)p(\lambda_1)$. Here we have explicitly used the \emph{measurement independence} assumption~\cite{Hall2010,Chaves2015} $p(\lambda_1 \vert x)=p(\lambda_1)$, that the measurement setting can be chosen independently of how the system is prepared. After the measurement, the system will be in a potentially different state $\Lambda_2$. For a projective measurement, $\Lambda_2$ is fully specified by the setting and outcome of the preceding measurement, and does not depend directly on the pre-measurement state $\Lambda_1$, that is $p(\lambda_2\vert x,a,\lambda_1)=p(\lambda_2\vert x,a)$. This implies that all correlations between $\Lambda_1$ and $\Lambda_2$ are mediated via the measurement.
To see this, note that the quantum probability distribution after a series of three sequential measurements is given by
\begin{equation}
\label{eq_quantum_temporal}
p_{\mathrm{Q}}(a,b,c \vert x,y,z)= p(c\vert b,y,z)p(b\vert a,x,y)p(a\vert x) \, ,
\end{equation}
where $p(c\vert b,y,z)=\tr \left[M^{z}_{c} \rho^{y}_{b} \right]$, $p(b\vert a,x,y)=\tr \left[M^{y}_{b} \rho^{x}_{a} \right]$ and $p(a\vert x)=\tr \left[M^{x}_{a} \rho_1 \right]$. In particular, any potential correlation between the measurement outcome $c$ in the third time step with the measurement setting and outcome in the first time step ($x$ and $a$, respectively), is screened-off by the intermediate measurement setting and outcome ($y$ and $b$), that is, $p(c\vert a,b,x,y,z)=p(c\vert b,y,z)$. This is similar to the no-signalling constraints arising in a Bell scenario and thus imposes restrictions to the classical causal models describing such a scenario. Specifically, a causal model that reproduces this independence without resorting to fine-tuning cannot contain a causal link of the form $\Lambda_2 \rightarrow \Lambda_3$, since such a link can generate unwanted correlations between the variables $X,A$ and $C$ (not mediated by $Y,B$). In fact, any model that contains such a link, can only satisfy the condition $p(c\vert a,b,x,y,z)=p(c\vert b,y,z)$ by virtue of causal parameters chosen from a set of measure zero~\cite{Pearlbook}, that is, the parameters are fine-tuned in such a way that these correlations are hidden from the observational data~\cite{Wood2015}.

Following the above description, the case of 3 sequential projective measurements on a single qubit can be represented in terms of the causal structure in Fig.~\ref{fig.dag1}. The temporal correlations $p(a,b,c \vert x,y,z)$ compatible with this causal structure can then be decomposed as (with straightforward generalization to more time steps)
\begin{align}
\label{eq.temporal_decomposition}
p(a,b,c \vert x,y,z) & = \sum_{\lambda_1,\lambda_2,\lambda_3} p(a \vert x,\lambda_1)p(b \vert y, \lambda_2) \\ \nonumber
& p(c \vert z,\lambda_3) p(\lambda_1) p(\lambda_2 \vert x,a) p(\lambda_3 \vert y,b) .
\end{align}
Note that without any restrictions on the dimensionality the hidden states $\Lambda_i$ can contain the full information about the measurement settings and outcomes of the previous time steps. This implies that, in contrast to the spacelike Bell scenario, all temporal correlations of the form of Eq.~\eqref{eq_quantum_temporal} can be faithfully reproduced by the classical causal model in Fig.~\ref{fig.dag1}.

This naturally raises the question whether further restrictions on the classical causal model, might reveal a quantum advantage. Similar to the so-called prepare-and-measure scenarios~\cite{Gallego2010}, one might expect that quantum systems of a given dimension give rise to measurement statistics that cannot be reproduced by classical systems $\Lambda_i$ of the same dimension. Since restriction on the dimension of $\Lambda_i$ in models of the form of Fig.~\ref{fig.dag1} would lead to non-convex sets that are technically very challenging to characterize~\cite{Chaves2016,Lee2015}, one typically considers convex relaxations. The resulting models contain the model of interest as a special case, but allow for shared randomness between the parties. For example, using the results of Ref.~\cite{Toner2003} we show in the Appendix~\ref{Sec:Supp1} that classical hidden states of dimension 4 (two bits of classical information), together with shared randomness between the parties, are enough to reproduce all correlations from a series of projective measurements on a single qubit. For a fair comparison to a qubit, we then considered hidden states of dimension two and could not find a difference between quantum and classical correlations. In light of these results it would be very interesting to test the model of Fig.~\ref{fig.dag1} with two-dimensional hidden states and no shared randomness. Due to the complexity of characterizing such non-convex sets, however, this remains an open question.

Besides restrictions on the dimension of the hidden state, certain physical constraints and assumptions might naturally lead to weaker causal models for a sequence of projective measurements on a single qubit. A well-known example is the Leggett-Garg (LG) model~\cite{Leggett1985} for testing macroscopic realism. This model is based on the assumption of noninvasive measurability, stating that it should be possible, in principle, to determine (measure) the state of a system without perturbing it. This is a special case of Eq.~\eqref{eq.temporal_decomposition}, where the hidden variable state is unchanged by the measurements and constant throughout the experiment, that is, $\lambda_i=\lambda$, which implies that
\begin{equation}
\label{eq.temporal_decomposition2}
\begin{split}
p(a,b,c \vert x,y,z) = \sum_{\lambda} & p(a \vert x,\lambda)p(b \vert y, \lambda)  \\
& p(c \vert z,\lambda) p(\lambda) .
\end{split}
\end{equation}
Note that this is the usual local hidden variable description encountered in Bell's theorem. Further, since the expectation values of a sequence of projective measurements on a single qubit are the same as for local measurements on a pair of entangled particles~\cite{Fritz2010}, quantum correlations also violate macroscopic realism, manifest as violations of so-called Leggett-Garg inequalities~\cite{Emary2013}. The result, however, relies critically on the assumption of noninvasive measurability, which is difficult to justify for a single quantum system. In fact, quite generally, the correlations obtained by sequential measurements on a quantum system will display signaling (e.g $p(b\vert x,y) \neq p(b\vert x^{\prime},y)$) as opposed to the model in Eq.~\eqref{eq.temporal_decomposition2} that only allow for non-signaling correlations. In other terms, the Leggett-Garg model implies independence relations that are not observed in the experiment. In this sense, in order to test incompatibility with the Leggett-Garg model we do not need to take the strength of the correlations into account and simply look for violations of independence relations implied by the model~\cite{Kofler2015}.

This raises the question of whether one can find examples of temporal quantum experiments to which classical faithful causal models can reproduce all independence relations while at the same time being incompatible with the generated correlations. Next we show that this is indeed the case.

\section{Quantum incompatibility with a weaker classical model for sequential projective measurements}
\begin{figure}[!t]
  \begin{center}
\includegraphics[width=0.9\columnwidth]{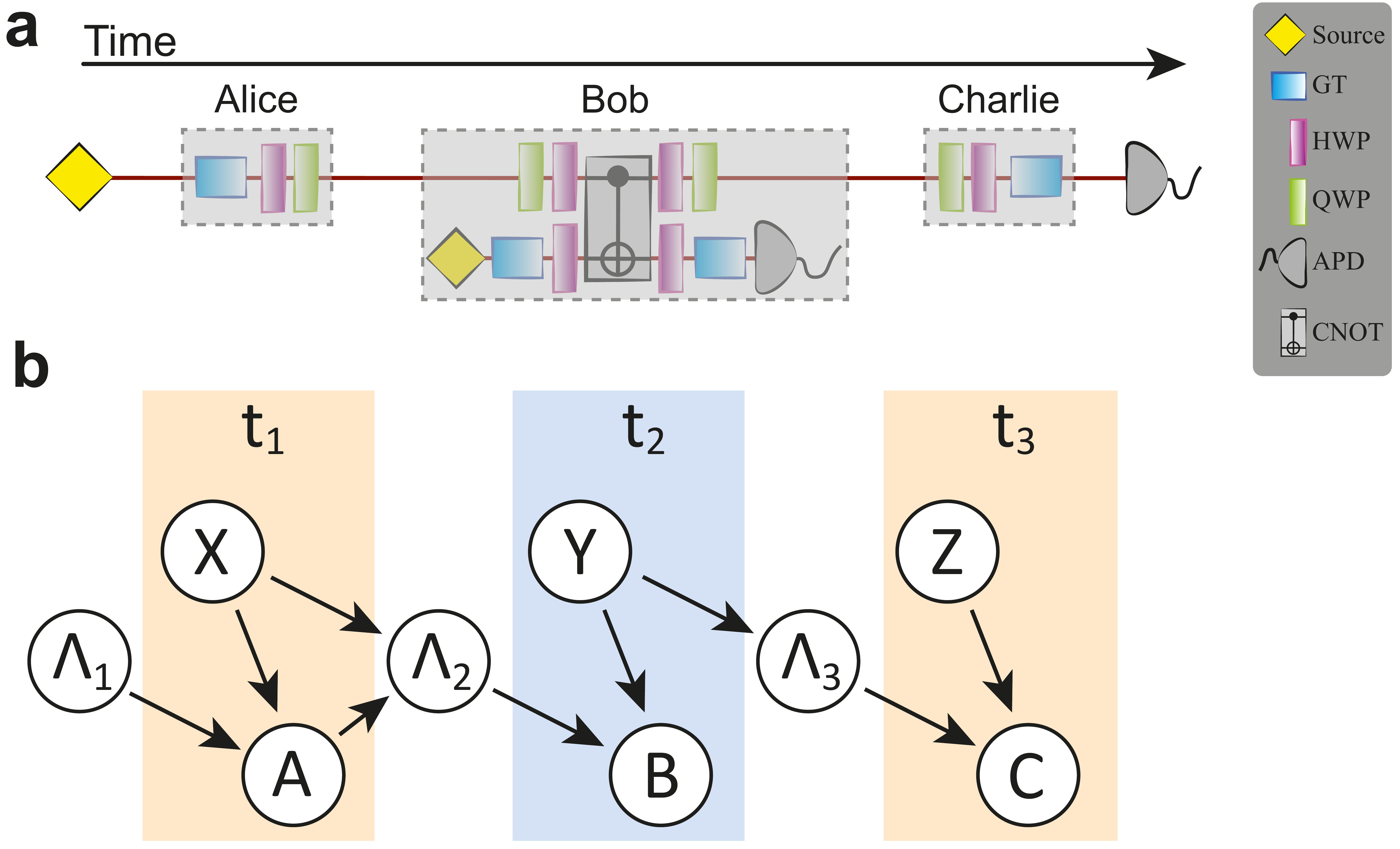}
  \end{center}
\vspace{-1em}
\caption{\textbf{The causal model of Eq.~\eqref{eq.temporal_decomposition3} for a sequence of projective measurements.} \textbf{(a)} Experimentally, the first and last parties, Alice and Charlie, measure the polarization of the photon directly, while the intermediate party, Bob, implements a non-destructive measurement by coupling the system to a meter photon using the CNOT gate in Fig.~\ref{fig:Setup}. Note that, this experiment naturally satisfy the independence $p(c\vert x)= p(c)$.
\textbf{(b)} The causal graph corresponding to Eq.~\eqref{eq.temporal_decomposition3} faithfully captures this independence. This model is slightly weaker than Fig.~\ref{fig.dag1}, since the ontic state $\Lambda_3$ depends only on the previous measurement setting $Y$, but not on the outcome $B$.}
\label{fig.dag1Weaker}
\end{figure}

From a causal inference perspective, given some observed probability distribution the goal is to find a faithful causal model reproducing all the (conditional) independence relations implied by this distribution. As a concrete example, consider a sequence of three projective measurements (see Fig.~\ref{fig.dag1}a) on an initial maximally mixed qubit state $\rho_1=\openone/2$. It is easy to verify that for arbitrary projective measurements, the measurement outcome of the third measurement is independent of the setting of the first measurement, i.e.\ $p(x,c)=p(x)p(c)$. In this case, the causal model in Fig.~\ref{fig.dag1}b is not faithful any longer because it allows for correlations between the variables $X$ and $C$. Instead, the most general causal model reproducing such independence relation is shown in Fig.~\ref{fig.dag1Weaker} where, in comparison with the causal model in Fig.~\ref{fig.dag1}, the causal link between the variable $B$ and $\Lambda_3$ is removed. Any distribution compatible with this model has a decomposition given by
\begin{align}
\label{eq.temporal_decomposition3}
p(a,b,c \vert x,y,z) & = \sum_{\lambda_1,\lambda_2,\lambda_3} p(a \vert x,\lambda_1)p(b \vert y, \lambda_2) \\ \nonumber
& p(c \vert z,\lambda_3) p(\lambda_1) p(\lambda_2 \vert a, x) p(\lambda_3 \vert y) .
\end{align}
Crucially, this classical model faithfully captures the observed independence $p(x,c)=p(x)p(c)$ that holds for a wide range of relevant experimental scenarios. This includes arbitrary measurements on an initially maximally mixed state, as well as arbitrary initial states in the \textsc{xy}-plane of the Bloch-sphere for the measurements in the experimental implementation below.

As detailed in Appendix~\ref{Sec:Supp1} any correlations compatible with Eq.~\eqref{eq.temporal_decomposition3} must respect the inequality
\begin{align}
\label{eq.tau3}
& S_{\tau_3}=\mean{A_0B_0}+\mean{A_0B_1}+\mean{A_1B_0}-\mean{A_1B_1}  \\ \nonumber
& -\mean{BC_{001}}-\mean{BC_{101}}-\mean{BC_{010}}-\mean{BC_{110}} \leq 6 ,
\end{align}
where the joint expectation values are defined as $\mean{A_xB_y}=\sum_{a,b}(-1)^{a+b} p(a,b \vert x,y)$ and $\mean{BC_{xyz}}=\sum_{a,b,c}(-1)^{b+c} p(a,b,c \vert x,y,z)$.
This inequality can, however, be violated by a sequence of projective measurements on any initial qubit state. For instance, choosing measurement settings $A_0=\hat Z$, $A_1=\hat X$, $B_0=-C_1=(\hat Z+\hat X)/\sqrt{2}$, and $B_1=-C_0=(\hat Z-\hat X)/\sqrt{2}$ (where $\hat X$ and $\hat Z$ are the Pauli operators) obtains a value of $S_{\tau_3}=4\sqrt{2} > 6$. Furthermore, for any initial state in the $\textsc{xy}$-plane of the Bloch sphere, the resulting probability distribution $p(a,b,c \vert x,y,z)$ respects the independence relation $p(c\vert x)= p(c)$ implied by the model under test. Through unitary rotations this implies that for any fixed initial quantum state, one can generate temporal correlations that cannot be explained by non-fine-tuned models.

\begin{figure}[ht!]
  \begin{center}
  \includegraphics[width=0.95\columnwidth]{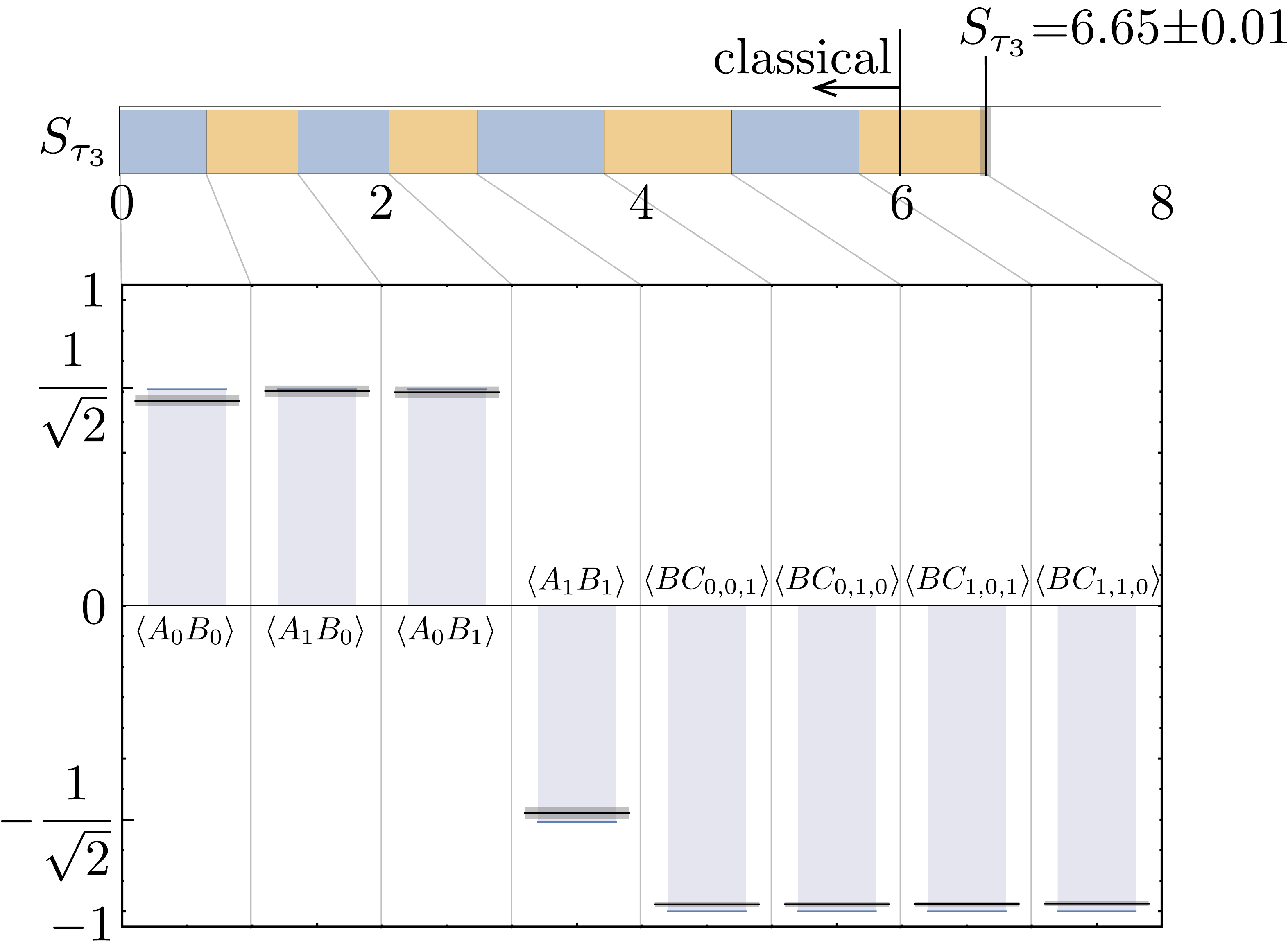}
  \end{center}
\vspace{-1.5em}
\caption{\textbf{Experimental results for testing Inequality~\eqref{eq.tau3} with a sequence of measurements on a single quantum systems}. The top part shows the experimental value of $S_{\tau_3}$ (vertical black bar) together with the $3\sigma$ statistical confidence region (grey shaded area). It also shows the individual contributions from the 8 expectation values (alternate blue and orange shadings), which are compared to their theoretical expectation in the bottom part of the figure. Here, blue shaded bars correspond to theoretical expectation values for ideal states and measurement and the horizontal black bars represent the experimental values with $3\sigma$ statistical error bars shown as grey shaded areas (almost within the black bars).}
  \label{fig:Results1}
\end{figure}

Experimentally we test inequality~\eqref{eq.tau3} with photonic polarization qubits for an initial maximally mixed state, see Fig.~\ref{fig.dag1Weaker}a. For the intermediate measurement the system is coupled to a meter in the state $\ket{0}$. A measurement of the meter in the computational basis $\{\ket 0,\ket 1\}$ achieves a projective measurement of the system in a basis that is chosen by appropriate single-qubit unitaries applied to the system before and after the interaction. Notably, this measurement design can be straightforwardly generalized to more than three parties by replicating the von Neumann measurement. The experimental results in Fig.~\ref{fig:Results1} demonstrate a clear violation of inequality~\eqref{eq.tau3} by a series of three projective measurements on a single qubit, achieving a value of $S_{\tau_3}=6.65^{+0.01}_{-0.01}$.

\section{Discussion}
Our results show, both in theory and experiment, that the discrepancy between the classical and quantum descriptions typically associated with spatial correlations extends to temporal correlation scenarios. The latter arise naturally in situations where spacelike separation cannot be practically guaranteed, or when a sequence of measurements is performed on the same quantum system. In the case of spatial correlations with an underlying time order, we have shown that a Bell-type inequality, designed to test classical models augmented with limited communication, displays a quantum violation if and only if the quantum protocol is also augmented with communication power. This highlights that a quantum advantage persists in a fair comparison of classical and quantum resources.

In a purely temporal-correlations scenario we have shown that, as opposed to spatial Bell scenarios~\cite{Wood2015}, there is a faithful (non-fine-tuned) classical model capable of simulating all correlations from projective measurements on quantum states. It remains an open question whether these models are equivalent when imposing the same dimensionality to quantum and classical models. Notwithstanding, we identified quantum violations of inequalities associated with a slightly less powerful classical model that includes the Leggett-Garg model as special case~\cite{Leggett1985}. We have experimentally observed such a quantum violation, which demonstrates a stronger form of non-classicality of the correlations arising from a temporal sequence of projective measurements. Furthermore, our experimental design allows for tuning of the intermediate measurement strength, which will enable future studies of non-Markovian models with some residual correlations between the ontic states $\Lambda_i$ over multiple time-steps~\cite{Ringbauer2015Superchannel}.

Our theoretical results are based on a causal modeling approach and thus formulated without reference to a background spacetime structure. The direction of time is only implicitly deduced from the flow of information between the parties or sequence of measurements. Experimentally, the considered scenarios are not subject to the locality loophole. However, the results rely on the related assumption that there is no hidden communication channel---other than the ones implied by the models in Figs.~\ref{fig.BellDag}b and~\ref{fig.dag1}---between the different time steps of the scenario under consideration. Practically, we also rely on a fair-sampling assumption to contend with imperfect detection efficiencies. 

Temporal correlation scenarios play an important role in communication complexity problems~\cite{Buhrman2010} and in the search for a physical principle behind quantum nonlocality, such as information causality~\cite{Pawlowski2009b}. Our results thus provide an avenue towards a more systematic understanding of the quantum advantage arising in such scenarios, that not only may lead to new ways of processing information but also to new insights into the nature of quantum correlations.

\begin{acknowledgments}
We thank C.~Budroni, F.~Costa, A.~Fedrizzi, and A.G.~White for helpful discussions and feedback, and T.~Vulpecula for experimental assistance. This work was supported in part by the Centres for Engineered Quantum Systems (CE110001013) and for Quantum Computation and Communication Technology (CE110001027), the Engineering and Physical Sciences Research Council (grant number EP/N002962/1), the Brazilian ministries MEC and MCTIC, the FQXi Fund, and the Templeton World Charity Foundation (TWCF 0064/AB38).
\end{acknowledgments}


%

\appendix
\renewcommand{\theequation}{A\arabic{equation}}
\renewcommand{\thefigure}{A\arabic{figure}}
\setcounter{equation}{0}
\setcounter{figure}{0}

\section{Classical models for a sequence of projective measurements on a qubit}
\label{Sec:Supp1}
\subsection{Classical simulation with hidden states of dimension 4} 
Note that the sequential measurement scenario can be mapped to an equivalent sequence of quantum teleportations~\cite{Bennett1993}. In the first time step Alice measures the state in her possession generating some local probability distribution $p(a \vert x)$, which she uses to prepare different states $\rho_a^x$ that she sends to Bob via the usual quantum teleportation protocol. Then Bob measures the teleported system generating a probability distribution $p(b \vert x,a,y)$ and prepares states $\rho_b^y$ that are teleported to Charlie. Clearly, any classical (hidden variable) protocol simulating the statistics of the measurements performed on the teleported states will immediately lead to a simulation of the equivalent sequence of projective measurements. As shown in Ref.~\cite{Toner2003}, this can be achieved using only two bits of classical communication between the parties and an arbitrary amount of shared randomness, implying that hidden states of dimension 4 and shared randomness are enough to obtain all quantum correlation obtained by a sequence of projective measurements on a single qubit.

For a fair comparison of classical and quantum resources we have also considered whether a classical message of dimension 2 (1 bit of communication) is enough to simulate all projective measurements on a qubit state. To that aim we considered a scenario with two time steps. If Alice has the choice between two measurements (i.e.\ $x=0,1$) then one classical bit can carry all the information about the input $X$ and, using the argument in the main text, all the one-way signalling correlations in this case can be simulated using a classical message of dimension 2. We thus considered the case were $X$ assumes at least 3 different values and fully characterized the set of classical correlations for dichotomic measurements ($a,b=0,1$) and with $x=0,1,2$ and $y=0,1$. The polytope corresponding to this scenario is described by 864 inequalities (many being equivalent under the allowed symmetries given by party, input and output permutations). Considering qubit states and arbitrary projective qubit measurements we could not find any violation of these inequalities.

It is interesting to note that among these inequalities we find the dimension-witness inequality from Ref.~\cite{Gallego2010}, given by
\begin{equation}
\mean{B_{00}} +\mean{B_{01}} +\mean{B_{10}} -\mean{B_{11}} -\mean{B_{20}}  \leq 3,
\end{equation}
where $\mean{B_{xy}}=\sum_{b=0,1} (-1)^{b}p(b \vert x,y)$. As shown in Ref.~\cite{Gallego2010}, this inequality can be violated by measurements on a qubit. There, however, a slightly different situation is considered, the so-called prepare-and-measure scenario, where the variable $X$ uniquely identifies the state $\rho_x$ being prepared and to be measured in the second time step. In our case the states $\rho^a_x$ to be measured in the second time step will depend on $X$, but also on the measurement outcome $A$, a feature that seems to be enough to preclude any violation of the inequality above (or any other defining the scenario). It would be interesting to derive inequalities for more general scenarios including more measurement settings or time steps to see whether any violations can be found.

\subsection{Derivation of Bell-type inequalities bounding classical models for a sequence of projective measurements}
\begin{figure}[!t]
  \begin{center}
\includegraphics[width=0.5\columnwidth]{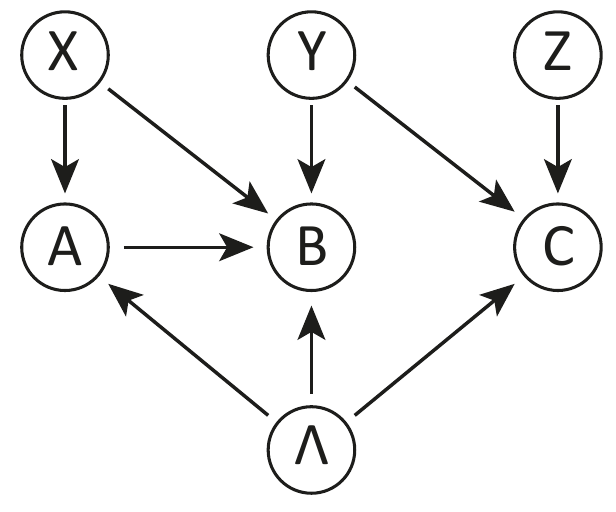}
  \end{center}
\vspace{-1em}
\caption{A DAG for the tripartite temporal correlations scenario that contains the DAG in Fig.~\ref{fig.dag1} of the main text as a special case. As shown in Ref.~\cite{Chaves2016causal} the correlations compatible with this DAG define a convex set such that Bell-type inequalities can be derived using standard convex optimization software~\cite{porta}.}
\label{fig.dag2}
\end{figure}

In order to derive Bell-type inequalities for the temporal scenario in Fig.~\ref{fig.dag1Weaker}, first note that all correlations compatible with such a model are also compatible with a model implying that
\begin{align}
\label{eq.scenario1}
p(a,b,c \vert x,y,z) =  & \sum_{\lambda} p(\lambda) p(a\vert x,\lambda) \\ \nonumber
& p(b\vert a,x,y,\lambda)p(c\vert y,z,\lambda).
\end{align}
This follows from the fact that the arrow between the hidden variable at a given time step and the next measurement outcome (e.g.\ $\Lambda_2 \rightarrow B$) can be replaced by directed arrows from the measurement choices (or measurement outcomes) of the previous step to the next one (e.g, $X \rightarrow B$ and $A \rightarrow B$) plus a local noise variable ($\Lambda_B \rightarrow B$). Note that these noise terms are implicitly present in the model in Fig.~\ref{fig.dag1Weaker}, where they have been absorbed into $\Lambda_1$, $\Lambda_2$, $\Lambda_3$. When making the above replacement, however, these local noise terms have to be introduced explicitly. They can then be combined into a variable $\Lambda=(\Lambda_A,\Lambda_B,\Lambda_C)$ that acts as a common ancestor and source of randomness for all the measurement outcomes, see Fig.~\ref{fig.dag2}. In principle, one would further have to impose the independence of the local noise terms, that is $p(\lambda_A,\lambda_B,\lambda_C)=p(\lambda_A)p(\lambda_B)p(\lambda_C)$. This, however, would define a non-convex set that is very difficult to characterize~\cite{Chaves2016,Lee2015}. Instead, we consider a more general convex relaxation of this set which contains the case of independent noise variables as a special case.

As detailed in Ref.~\cite{Chaves2016causal}, the probability distributions compatible with Eq.~\eqref{eq.scenario1} define a convex polytope that can be characterized in terms of finitely many extremal points. Given the list of extremal points one can resort to standard convex optimization software~\cite{porta} to find the dual description in terms of linear (Bell-type) inequalities.
For quantum correlations arising from a sequence of three projective measurements it follows that $\mean{ABC}= \mean{A}\mean{BC}$.
In other words, the full (three-point) correlator does not carry any information that is not already contained in the bipartite and single correlators. For this reason we focus our attention on inequalities involving the expectation values $\mean{AB}$ and $\mean{BC}$, which, due to the structure of the problem, can be defined in different ways. In the following we use
\begin{align}
\mean{AB_{x,y}}&=\sum_{a,b}(-1)^{a+b} p(a,b \vert x,y) \\
\mean{BC_{x,y,z}}&=\sum_{a,b,c}(-1)^{b+c} p(a,b,c \vert x,y,z) .
\end{align}
Since there is a causal link between $X$ and $B$, the correlations between $B$ and $C$ can explicitly depend on $X$. We then compute the Bell-type inequalities in this subspace, one of which is inequality~\eqref{eq.tau3} in the main text.

\subsection{Quantum violation of inequality~\eqref{eq.tau3}}
In order to search for a possible quantum violation of inequality~\eqref{eq.tau3}, we consider a single qubit in the initial pure state
\begin{equation}
\ket{\Psi}= \cos{\theta_0}\ket{0}+e^{i\phi_0}\sin{\theta_0}\ket{1} .
\end{equation}
At each time step we measure observables $\hat O_i$, parametrized by $\theta_i$ and $\phi_i$
\begin{equation}
\hat O_i= \cos{\phi_i}\sin{\theta_i}\hat X+\sin{\phi_i}\sin{\theta_i}\hat Y+\cos{\theta_i}\hat Z .
\end{equation}
The full correlator can then be expressed, as $\mean{ABC}= \mean{A}\mean{BC}$ with $\mean{A}=\cos{\theta_0}\cos{\theta_1}+\cos{(\phi_0-\phi_1)} \sin{\theta_0}\sin{\theta_1}$ and $\mean{BC}=\cos{\theta_2}\cos{\theta_3}+\cos{(\phi_2-\phi_3)} \sin{\theta_2}\sin{\theta_3}$. Hence, as expected, the projective measurement at the second time step (corresponding to the outcome $b$) destroys any correlations between the first and third time steps (outcomes $a$ and $c$, respectively). Similar to $\mean{BC}$, the correlations between the first and second time steps are given by $\mean{AB}=\cos{\theta_1}\cos{\theta_2}+\cos{(\phi_1-\phi_2)} \sin{\theta_1}\sin{\theta_2}$.

For projective measurements on a qubit state we thus obtain $\mean{BC_{x=0,y,z}}=\mean{BC_{x=1,y,z}}$. Since the first four terms of inequality~\eqref{eq.tau3} are nothing else than CHSH, this part can be maximized using the standard settings: $A_0=\hat Z$, $A_1=\hat X$, $B_0=(\hat X+\hat Z)/\sqrt{2}$, $B_1=(-\hat X+\hat Z)/\sqrt{2}$. The last four terms are maximized by setting $C_0=-B_1$ and $C_1=-B_0$, which obtains a value of $4$. In summary, quantum correlations can obtain $S_{\tau_3}=4+2\sqrt{2} >6$, thus violating inequality~\eqref{eq.tau3}. Curiously, this holds for any initial single-qubit state.

\section{Experimental details}
\label{Sec:Supp2}
To test inequality~\eqref{eq.M332} in the main text we prepare a single-parameter family of two-qubit states of the form
\begin{equation}
\begin{split}
\ket{\Psi}=\frac{1}{2}\bigl(&\sqrt{1+\kappa^2} (\ket{HH}+\ket{VV}) \\
&+ \sqrt{1-\kappa^2} (\ket{HV}+\ket{VH})  \bigr) .
\label{eq:SIStates}
\end{split}
\end{equation}
We generate this family of states by subjecting the separable state $\ket{D}\otimes(\sqrt{1+\kappa}\ket{H}+\sqrt{1-\kappa}\ket{V})/\sqrt{2}$ to a controlled-NOT gate, as shown in Fig.~\ref{fig:Setup} in the main text. The parameter $0\leq \kappa\leq 1$ turns out to be equal to the concurrence $\mathcal{C}$ of the resulting bipartite state, such that the generated amount of entanglement can be easily controlled by the initial state of the meter photon.

Using the fixed measurement protocol in the main text, the states of Eq.~\eqref{eq:SIStates} achieve a value of $S_{\text{1bit}}=4+(1+\kappa)\sqrt{2}$ in inequality~\eqref{eq.M332}. It is, however, possible to optimize the measurement settings such that ideally every pure entangled quantum state violates inequality~\eqref{eq.M332}. Specifically, this amounts to modifying the protocol in the main text such that for $m=0$ Bob measures $B_0=B_1=B_2=\hat X$, while for $x=1$ he measures $B_0=(\hat X+\kappa\hat Z)/\sqrt{1+\kappa^2}$, $B_1=(\hat X - \kappa\hat Z)/\sqrt{1+\kappa^2}$ and $B_2= - \hat X$. Using this protocol, states of the form of Eq.~\eqref{eq:SIStates} achieve $S_{\text{1bit}}=4+2\sqrt{1+\kappa^2}\geq 6$.

This is reminiscent of the observation that, using optimized measurement settings, the CHSH inequality \cite{Clauser1969} can be violated by every pure two-qubit entangled quantum state \cite{Horodecki1995}. In fact, inequality~\eqref{eq.M332} contains a CHSH inequality for the settings $A_1,A_2,B_0,B_1$, which can be violated in the usual way, while the additional communication in our protocol can be used to maximize the remaining four terms up to the maximal value of $4$ for every state of the form Eq.~\eqref{eq:SIStates}. Hence, the expectation value of inequality~\eqref{eq.M332} can be written as $S_{\text{1bit}}=4+S_{\textsc{chsh}}$, where $S_{\textsc{chsh}}$ is the CHSH parameter corresponding to the first four terms of inequality~\eqref{eq.M332}.

\end{document}